\documentclass[prd,twocolumn,showpacs,preprintnumbers,nofootinbib,amsmath,amssymb,nobalancelastpage,superscriptaddress]{revtex4}
\usepackage{graphicx}
\usepackage{bm}
\usepackage{dcolumn}
\usepackage{amsmath}
\usepackage{amssymb}
\usepackage{color}
\usepackage{url}
\usepackage{hyperref}

\usepackage{aas_macros}
\usepackage[normalem]{ulem}

\newcommand{\Msun}{M_\odot}
\newcommand{\mbh}{m_{\rm{BH}}}
\newcommand{\fDM}{f_{\rm{DM}}}

\newcommand{\erf}{{\rm{erf}}}
\newcommand{\mBH}{m_{\rm{BH}}}

\usepackage{color}

\begin{document}

\title{Dynamics of dwarf galaxies disfavor stellar-mass black holes as dark matter}

\author{Savvas M. Koushiappas}
\email{koushiappas@brown.edu}
\affiliation{Department of Physics, Brown University,  182 Hope St., Providence, RI 02912, USA}
\affiliation{Institute for Theory and Computation, Harvard University, 60 Garden Street, Cambridge, MA, 02138, USA}

\author{Abraham Loeb}
\email{loeb@cfa.harvard.edu}
\affiliation{Institute for Theory and Computation, Harvard University, 60 Garden Street, Cambridge, MA, 02138, USA}

\date{\today}

\begin{abstract}
We study the effects of black hole dark matter on the dynamical evolution of  stars in dwarf galaxies. We find that mass segregation leads to a depletion of stars in the center of dwarf galaxies and the appearance of a ring in the projected stellar surface density profile. Using Segue 1 as an example  we show that current observations of the projected surface stellar density rule out at the 99.9\% confidence level the possibility that more than 6\% of the dark matter is composed of black holes with a mass of few tens of solar masses. 
\end{abstract}

\pacs{95.35.+d, 04.25.dg, 95.85.Sz, 98.56.Wm}

\maketitle

The nature of dark matter remains an open question almost a century after its discovery \cite{1933AcHPh...6..110Z,1937ApJ....86..217Z}. Direct and indirect detection experimental searches \cite{2017PhRvL118b1303A,2017ApJ...834..110A,2015PhRvD..91h3535G} as well as the Large Hadron Collider \cite{2017arXiv170301651C,Aaboud:2016qgg} have been searching for a Weakly Interacting Massive Particle (WIMP) as a dark matter candidate with no positive results to date. The parameter space of axion dark matter is also shrinking with no evidence of a detection \cite{2016arXiv161208296S}. From the astrophysical perspective,  MAssive Compact Halo Objects (MACHOs) have for the most part been ruled out with microlensing experiments \cite{1986ApJ...304....1P,2001ApJ...550L.169A,2007A&A...469..387T}. At the high mass end, wide binaries in the Milky Way provide the strongest constraints \cite{2004ApJ...601..311Y,2009MNRAS.396L..11Q,2014ApJ...790..159M}.

An alternative to particle dark matter is that dark matter is composed of primordial black holes formed in the early universe prior to Big Bang Nucleosynthesis \cite{1974MNRAS.168..399C,1974A&A....37..225M,1975ApJ...201....1C,1976ApJ...206....8C}. These black holes can span a wide range of masses from $10^{-18}\Msun$ (where Hawking radiation \cite{1976PhRvD..13..191H} limits their current abundance) to $10^6 \Msun$. Recently Cosmic Microwave Background (CMB) constraints \cite{2016arXiv161205644A,2016arXiv161206811A} have ruled out primordial black holes with mass  $\sim10^2\Msun$ as the dominant form of dark matter.

The excitement surrounding the recent discovery of gravitational waves by LIGO \cite{2016PhRvL.116f1102A} led  to the suggestion that the observed black hole pairs that gave rise to the gravitational wave events (with a mass $m \sim 30 \Msun$) were primordial black holes \cite{2016PhRvL.116t1301B,2016PhRvD..94h4013C,2016PhRvD..94h3504C,2016PhRvD..94f3530G,2016arXiv161110069A}. It was shown that if the dark matter is composed of primordial black holes, then the LIGO events can be due to their mergers \cite{2016PhRvL.116t1301B}. The related mass range is weakly constrained by studies that probe the low mass end of black hole masses (e.g., microlensing) or studies that place constraints on the high-mass end (e.g., the CMB \cite{2016arXiv161205644A}, the half-light radius of dwarf galaxies  \cite{2016ApJ...824L..31B,2017ApJ...838....8L} and wide binaries in the Milky Way \cite{ 2014ApJ...790..159M}). 

In this {\it Letter} we examine this hypothesis in the context of the observed distribution of stars in dwarf galaxies. 
These are dark matter dominated galaxies, composed of old stars (e.g., \cite{2014ApJ...786...74F}) and located at distances of at least tens of parsecs to hundreds of kiloparsecs \cite{2012AJ....144....4M}. The number of  known systems of this type has increased over the last 10 years due to the Sloan Digital Sky Survey \cite{willman05a,zucker06a,belokurov07,belokurov08,2006ApJ...653L..29S} and the Dark Energy Survey~\cite{2015ApJ...805..130K,2015ApJ...807...50B}. 

A particular system that has been extensively studied over the past decade is the Segue 1 dwarf galaxy \cite{simon11,geha09,2011ApJ...738...55M}. Spectroscopic studies show that it is dark matter dominated \cite{simon07} and that its stellar population is old \cite{2014ApJ...786...74F}, with no evidence
of any major disruption or interaction \cite{2016ApJ...818...80W}. We use Segue 1 to demonstrate the effect of primordial black hole dark matter because it is well-studied, although a similar analysis can be applied to other dark matter dominated systems in the future.

Assuming that massive black holes are the dark matter (or some fraction $\fDM$ of it),  dwarf galaxies are collisionless systems with stars of mass $m_s \sim 1 M_\odot$ and black holes of mass $\mBH \gg m_s$. Both, stars and black holes respond to the underlying gravitational potential. 

The dynamics of a two component collisionless systems have been studied by Spitzer \cite{1940MNRAS.100..396S,1969ApJ...158L.139S}  who showed that relaxation leads to equipartition, where the average kinetic energy of the light component (e.g., stars) is  equal to the average kinetic energy of the heavy component (e.g., black holes).  Mass segregation takes effect over the relaxation timescale, whereas the light particles move outwards while the heavy particles sink towards the center. The physics of mass segregation is similar to dynamical friction where multiple scattering encounters between the two populations leads to energy exchange  (see e.g., \cite{2008gady.book.....B}). It follows naturally that the light particles move on average faster than the heavy particles and thus reside at larger radii. 

\begin{figure*}
\includegraphics[scale=0.46]{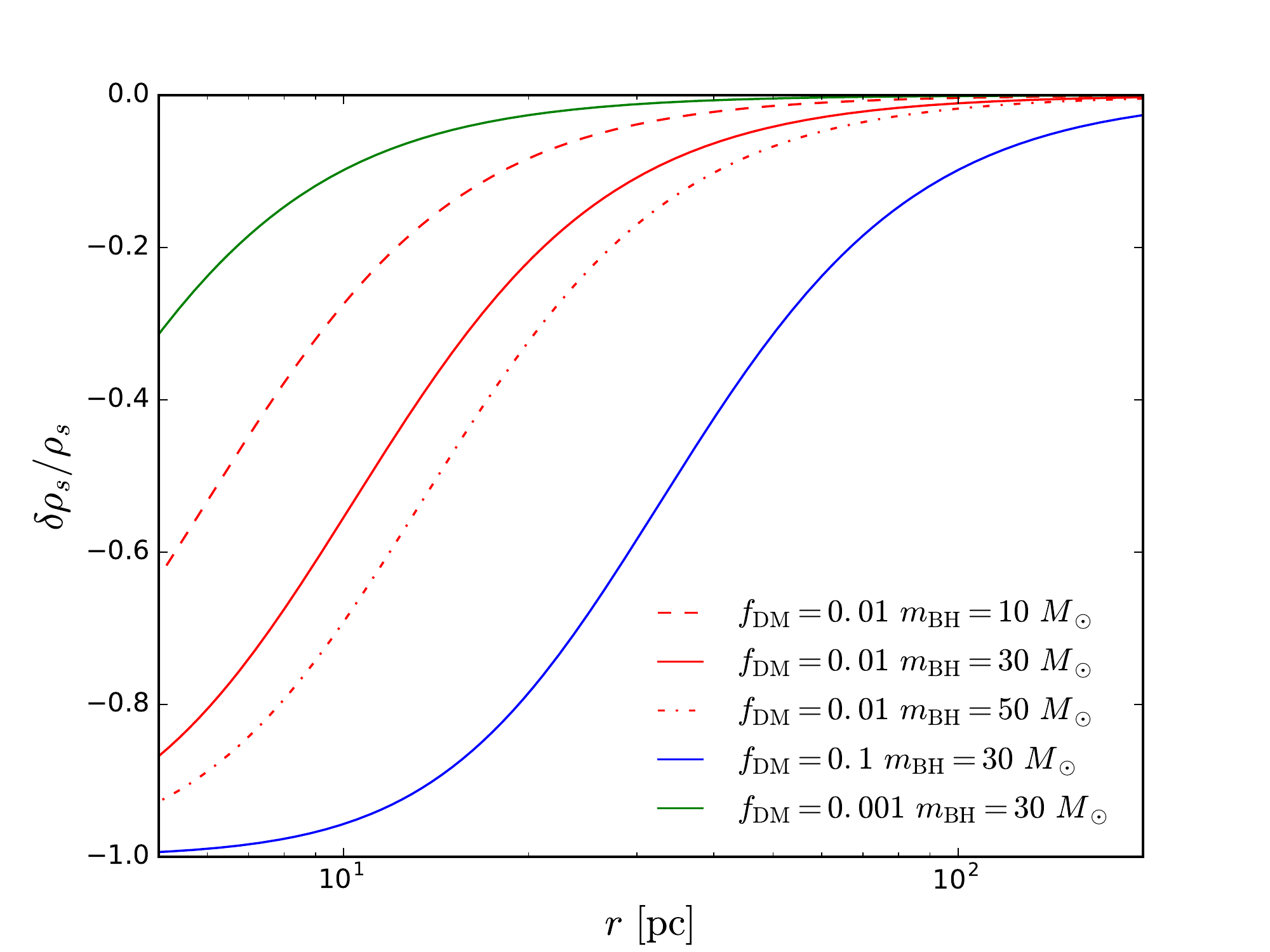}\includegraphics[scale=0.46]{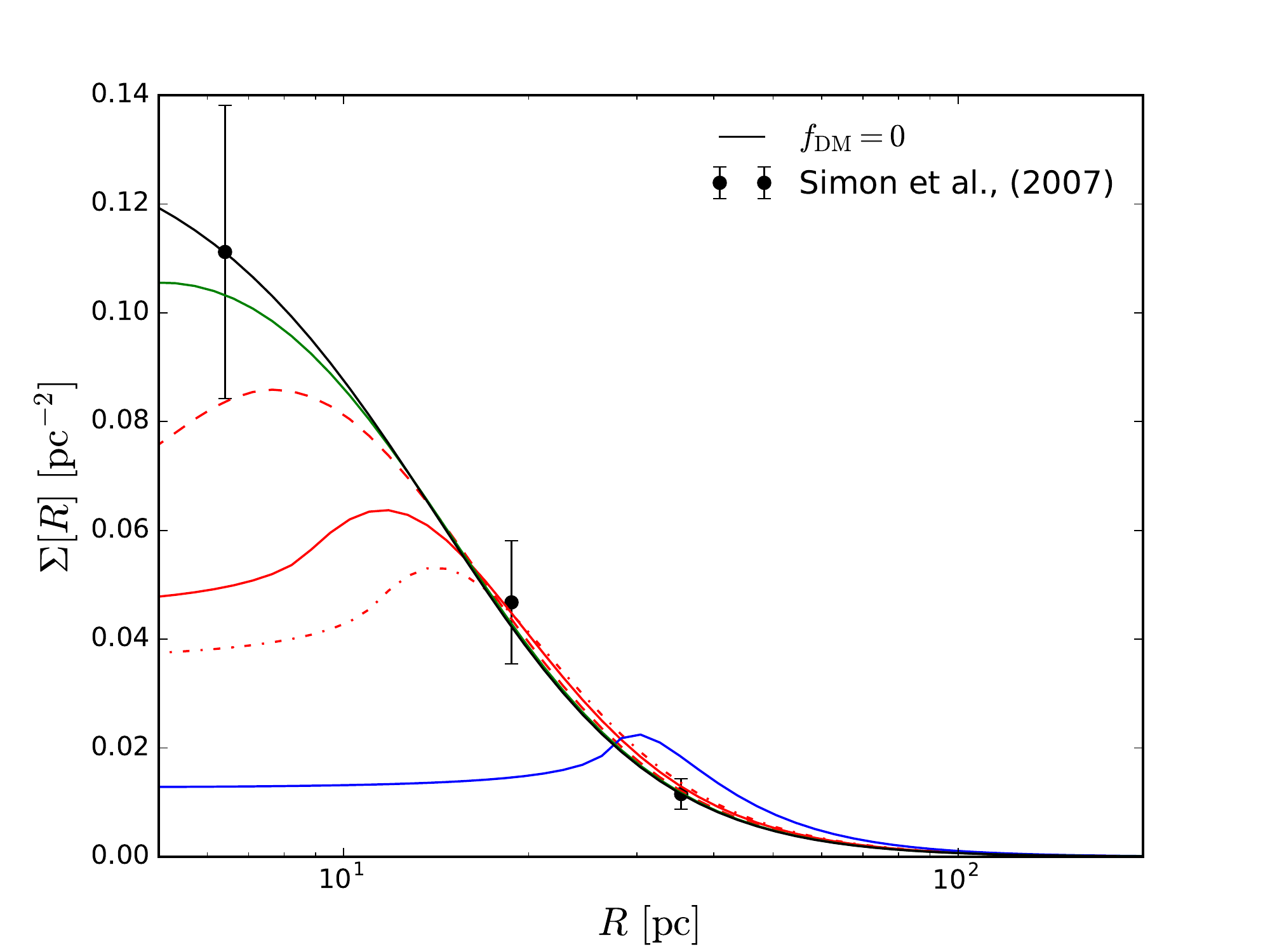}
\caption{\label{fig:fig1} {\it{Left}}: The evolved stellar deficit as a function of radius in Segue 1 for various fractions $f_{\rm{DM}}$ of black hole dark matter and black hole masses $m_{\rm{BH}}$. The deficit increases as $f_{\rm{DM}}$  and $m_{\rm{BH}}$ increase. {\it{Right}}: Projected stellar surface density of Segue 1. Data points represent the observed surface density \cite{simon11}. Black curve shows the case with no black hole dark matter. Line types and colors correspond to the same choices as in the left panel.}
\end{figure*}

We use these  results to explore the evolution of the stellar distribution in dwarf galaxies. We begin by defining the mean change in velocity due to scattering along the tangential and normal to the direction of motion of the star as  $ \Delta v_\parallel $ and $ \Delta v_\perp $ respectively. Assuming that both species (stars of mass $m_s$ and black holes of mass $\mbh$) are described by a Maxwellian velocity distribution function, the diffusion coefficient (average change of kinetic energy per unit mass and time) of stellar particles  due to  their scattering off black holes is \cite{2013degn.book.....M} 
\begin{eqnarray} 
\label{eq:deltaE}
\langle \Delta E \rangle_s &=& v_s \langle \Delta v_{s,\parallel} \rangle + \frac{1}{2} \langle ( \Delta v_{s,\parallel})^2 \rangle + \frac{1}{2} \langle ( \Delta v_{s,\perp} )^2 \rangle  \\
&=& \frac{ 4  \pi G^2 \, \mBH  \, \rho_{\rm{BH}} \, \ln \Lambda}{ v_s} \nonumber  \\ 
&\times& \left[  - \frac{m_s}{\mBH} \erf (X) + \left( 1 + \frac{m_s}{\mBH} \right) X \, \erf '(X) \right], \nonumber  
\end{eqnarray}
where prime denotes a derivative with respect to $X$, $X \equiv v / \sqrt{2} \sigma_{\rm{BH}}$, $\sigma^2_{\rm{BH}} = \langle v^2_{\rm{BH}} \rangle$,  $\ln \Lambda \approx 10$ is the Coulomb logarithm, and $G$ is the gravitational constant. 
The mean change of  kinetic energy of the stars $E_s =   m_s \langle {v_s}^2 \rangle /2$ is 
\begin{equation} 
\frac{dE_s}{dt}  = \sqrt{\frac{2}{\pi}} \frac{1}{\sigma_s^3} \int_0^\infty m_s \langle \Delta E \rangle_s \, v_s^2 e^{-v_s^2 / 2 \sigma_s^2} dv_s.
\label{eq:dKEdt}
\end{equation}
Substituting Eq.~(\ref{eq:deltaE}) in Eq.~(\ref{eq:dKEdt}) and integrating by parts we get \cite{2013degn.book.....M}, 
\begin{equation} 
\frac{dE_s}{dt}   = \frac{\sqrt{96 \pi} G^2 m_s \rho_{\rm{BH}} \, \ln \Lambda}{[ \langle v_s^2 \rangle + \langle v_{\rm{BH}}^2 \rangle ]^{3/2}} \left[ \mBH \langle v_{\rm{BH}}^2 \rangle - m_s \langle v_s^2 \rangle \right]. 
\label{eq:dKEdt2}
\end{equation}
Equation~(\ref{eq:dKEdt2}) shows that when $\mBH \langle v_{\rm{BH}}^2 \rangle = m_s \langle v_s^2 \rangle $ there is no energy exchange between the two populations. If  $\langle v_{\rm{BH}}^2 \rangle  \approx \langle v_s^2 \rangle \equiv \sigma^2$,  the timescale for stars and black holes to reach  equipartition is $t_{\rm{relax}} = E_s/(dE_s / dt)$ which based on the virial theorem can be written as $t_r \approx (N / 8 \ln N ) \tau_c$, where $\tau_c = r / \sigma$ is the crossing time and $N$ is the number of particles. If the system is dominated by black holes (as is the case here), then stars will reach equipartition soon as the black holes establish a collisional steady state. 

For Segue 1,  $\sigma = 3.7^{+1.4}_{-1.1} {\rm{km \, s^{-1}}}$, the half light radius is $29^{+8}_{-5}$ pc, and the mass within half light radius is $5.8^{+8.2}_{-3.1} \times 10^5$ \cite{simon11,2012AJ....144....4M}. Assuming that 10\% of dark matter is in black holes of mass $\mBH = 30\Msun$, the ratio of relaxation time to Hubble time is $\sim 0.01$. Thus, mass segregation and equipartition must have already taken place in Segue 1 by the present epoch\footnote{The quoted relaxation time is directly proportional to the fraction of dark matter in black holes. If for example the fraction of dark matter is $100\%$ ($1\%$) the ratio of relaxation time to Hubble time is $\sim 0.1$ ($\sim 0.001$).}. Other dwarf galaxies with similar relaxation times are Bootes II, Segue II, Wilman 1, Coma and Canes Venatici II. All other known dwarf galaxies have relaxation times that are at least a factor of 10 higher. 

We proceed by assuming that the initial distribution of stars is described by a Plummer profile. This is justified for two reasons: first, Plummer profiles are known to be acceptable fits to the present-day distribution of stars in dwarf galaxies, and second, a Plummer profile has an inner core. Anything steeper than a cored profile such as Plummer will exhibit even more severe effects of mass segregation\footnote{An exponential profile can also be used (see \cite{martin08}), with similar results.}.

We follow  \citet{2016ApJ...824L..31B} and calculate the evolution of radial shells by using the virial theorem and the diffusion coefficient for weak scattering of stars off black holes (see also \cite{2008gady.book.....B}). The differential equation that governs the evolution of radial mass shells as a function of time is then 
\begin{equation} 
\frac{dr}{dt} = \frac{ 4 \sqrt{2} \pi \, G \, f_{\rm{DM}} \mbh}{\sigma} \ln \Lambda \left( \alpha \frac{M_s}{\rho_{\rm{DM}} \, r^2} + 2 \, \beta \, r \right)^{-1}. 
\label{eq:drdt}
\end{equation}
We adopt for Segue 1 $\alpha = 0.4$, $\beta = 10$ (see \citet{2016ApJ...824L..31B}) and a total mass in stars of $M_s = 340 \Msun$ \cite{2012AJ....144....4M}. The choice of values for $\alpha$ and $\beta$ is such that the effects of mass segregation are minimal and thus provide a conservative choice (the result is insensitive to the choice of $\alpha$ as the density of stars is much less than the density of dark matter; lower values of $\beta$ simply imply a higher normalization of the $r \sim t^{1/2}$ solution to Eq.~(\ref{eq:drdt})). 

The stars are initially distributed in a Plummer profile with a scale radius of $r_s = 16 {\rm{pc}}$. This value is 25\% smaller than the currently measured value of the Plummer scale radius. Any other choice would lead to stronger constraints on black hole dark matter\footnote{We confirmed this assumption by repeating the analysis for a suite of initial scale radii of a Plummer profile as well as by assuming an isothermal sphere or a Hernquist profile as the initial distribution. All these options led to stronger constraints to black hole dark matter.}.  We assume that the dark matter distribution is described by a generalized NFW profile \cite{1996ApJ...462..563N} , whose parameters $\alpha$, $\beta$, $\gamma$, $\rho_s$ and $r_s$ as defined in Equation (7) of \cite{2015ApJ...801...74G} are given by the median values obtained by the MCMC analysis of \citet{2015ApJ...801...74G} . The median value of the profile parameters does not necessarily correspond to the median value of the density at all radii. We repeated the calculation by assuming the median of the density at each integrated radial shell and find that the deviations are negligible. In addition, repeating the calculation by marginalizing over all the kinematically-allowed distributions of dark matter also has negligible effects on the results.

We assume that at $t=0$ the outer envelope of the profile is similar to that observed at the present epoch. Any evolution of the stellar density profile should leave the outer regions of the stellar population unaffected. Given that at present the half light radius of Segue 1 $\sim 20$ pc, we set the profile to zero at a reasonably large radius of 300 pc.

We integrate Eq.~(\ref{eq:drdt}) over 12 Gyrs to obtain the evolution of each radial shell as a function of time. We find two main effects  of black hole dark matter. First, each initial radial distance (with stars interior to it)  moves outwards, with the displacement decreasing as the radius increases. There is no shell crossing and as stars in the outer regions remain unaffected, we find that stars that were displaced by black holes  lead to the presence of a spherical shell overdensity. The depletion of stars in the inner regions leads to the {\it prediction of a stellar ring in projection}\footnote{ Note that we ignore the effects of evaporation for two reasons. First, the evaporation timescale is $\sim {\cal{O}}(10-100)$ longer than the relaxation timescale and thus mass segregation will take place well before any effects of evaporation appear. Second, evaporation would deplete stars from the inner regions and therefore  augment the effects we observe here.  }.

The left panel of Figure~\ref{fig:fig1} shows the present-day evolution change of the stellar deficit, $\delta \rho_s / \rho_s \sim [r(0)/r(t)]^3 -1$ as a function of radius. Increasing the fraction of black hole dark matter leads to a larger depletion of stars in the center of the galaxy. A similar effect is obtained when the fraction of black hole dark matter is fixed but the black hole mass increases. The right panel of Figure~\ref{fig:fig1} shows the projected stellar surface density profile compared to the observed stellar profile density obtained from the stars identified in \citet{simon11} , binned in radii of equal number of stars (with Poisson errors).

\begin{figure}
\includegraphics[scale=0.47]{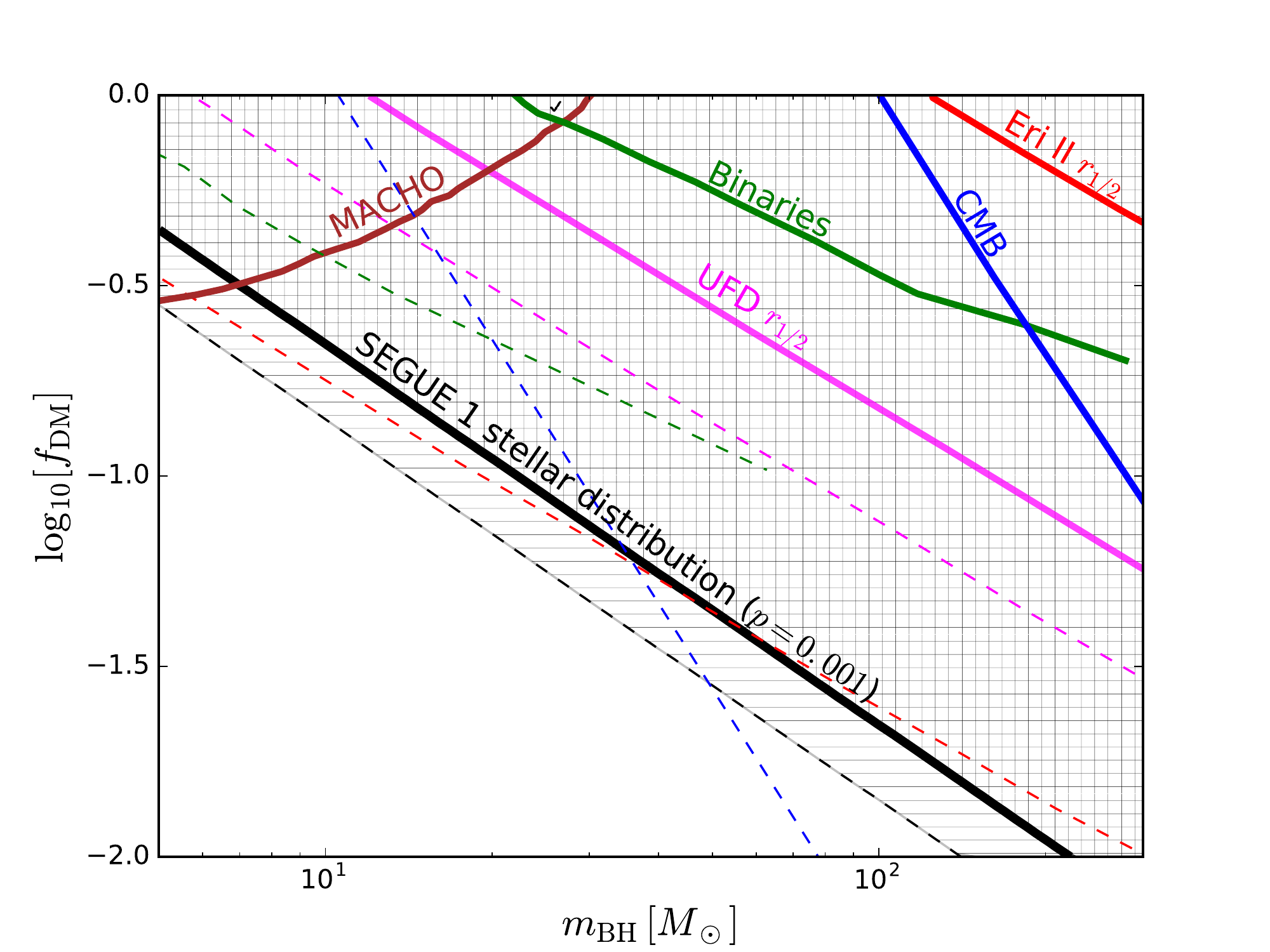}
\caption{\label{fig:chi2}  Constraints from the distribution of stars in Segue 1 on the fraction of dark matter in the form of black holes, $f_{\rm{DM}}$,  as a function of black hole mass $m_{\rm{BH}}$.  The solid (dashed) black contour corresponds to a $p$-value of 0.001 for the most (least) conservative case where the velocity dispersion of Segue 1 is 4.1 ${\rm{km}} \, {\rm{s}}^{-1}$ (2.7 ${\rm{km}} \, {\rm{s}}^{-1}$ ). We also show limits from the evolution of the half light radius of  the Eridanus II dwarf galaxy as well as other ultra faint dwarfs\ (UFDs) \cite{2016ApJ...824L..31B}, Milky Way wide binaries (using the 25 most halo like binaries) \cite{ 2014ApJ...790..159M}, microlensing limits from Eros-2 \cite{2007A&A...469..387T} and MACHO experiments \cite{2001ApJ...550L.169A}, and constraints from CMB photoionization from accretion onto primordial black holes \cite{2016arXiv161205644A}.  In all these cases, the solid lines correspond to the most conservative choice of parameters in these calculations while the thin dashed lines correspond to the least conservative choices. The stellar distribution in Segue 1 improves constraints for masses greater than   $6 \Msun$. 
}
\end{figure}

We use the observed distribution of stars to place constraints on the evolved light profile when there is a non-zero fraction of black hole dark matter. For each assumed value of $f_{\rm{DM}}$ and $\mBH$, we compute the evolved projected stellar surface density profile and compare it with the observed stellar profile \cite{simon11}. We assign a $\chi^2$ test statistic to each choice of $f_{\rm{DM}}$ and $\mBH$ and compute the corresponding $p-$value for 3 degrees of freedom. The result is shown in Figure~\ref{fig:chi2}. Black hole fractions greater than  6\% (20\%) for $\mBH = 30 \Msun$ ($\mBH = 10 \Msun$) are ruled out at the 99.9\% confidence level. Figure~\ref{fig:chi2} compares our results to previous constraints from  the observed half-light radius of the Eridanus II dwarf galaxy \cite{2016ApJ...824L..31B}, microlensing studies \cite{2001ApJ...550L.169A,2007A&A...469..387T}, CMB photoionization limits from accretion onto primordial black holes \cite{2016arXiv161205644A} and constraints from wide binaries in the Milky Way \cite{ 2014ApJ...790..159M}. The light profile of Segue 1 improves constraints on masses greater than   $ 6 \Msun$.

%

\begin{figure*}
\includegraphics[scale=0.4]{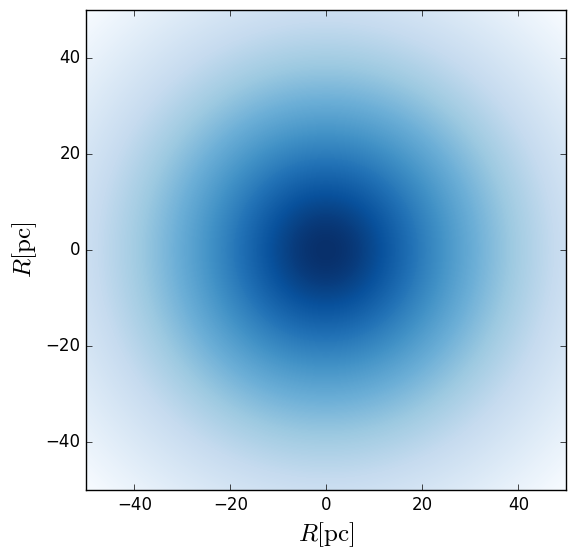} \includegraphics[scale=0.4]{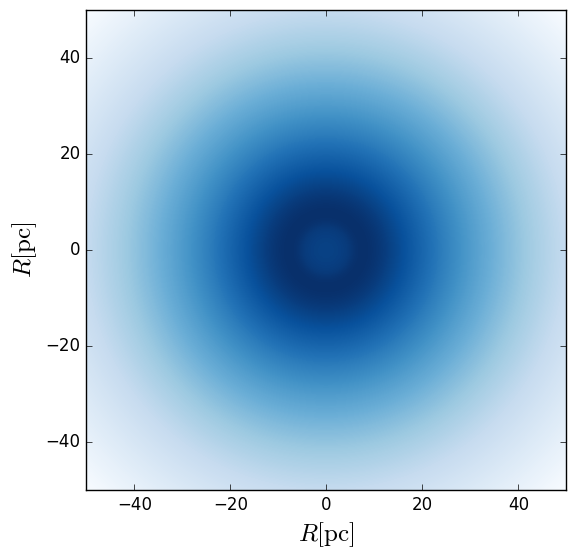}\includegraphics[scale=0.4]{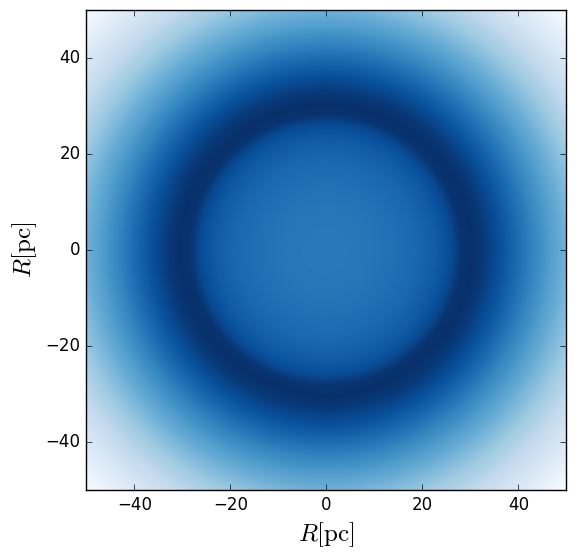}
\caption{\label{fig:simulations} Simulated effects of mass segregation in Segue 1. {\it Left}: projected stellar mass density in the case where there is no black hole dark matter and the dark matter distribution is smooth. {\it Middle}:  Similar to left, with 1\% of the dark matter in 10 $\Msun$ black holes. Mass segregation leads to the depletion of stars at the center and the presence of a ring in the projected stellar density. In both cases (left and middle panels), the half-light  radius is within the observed range of $29^{+8}_{-5}$ pc. With the current amount and quality of data it is not possible to distinguish between these two cases; however, future observations may be able to constrain such models with an increase in the number of observed member stars. {\it Right}: An  example where 30 $\Msun$ black holes constitute 10\% of the dark matter density. This scenario is ruled out by current observations.}
\end{figure*}

The above constraints can be improved if future observations would reveal more stars in Segue 1 (as well as other dwarf galaxies). Figure~\ref{fig:simulations} show a simulated smoothed projected stellar density of Segue 1 in the case where there is no black hole dark matter present (left panel) and when 1\% of dark matter is in 10 $\Msun$ black holes (middle panel). Mass segregation depletes the core, however with current observations $f_{\rm{DM}}=1\%$ in $m_{\rm{BH}}=10 \Msun$ is still allowed (see Figure~\ref{fig:chi2}). The half-light radius in both cases is within the error of the currently assumed half-light radius of Segue 1, so in the absence of any additional information it is impossible to distinguish between the two cases. For comparison, the right panel of Figure~\ref{fig:simulations} depicts the projected surface density profile at the currently excluded case where 10\% of the dark matter is in black holes of mass $\mBH = 30 \Msun$. 

A future improvement to our analysis could involve a Fokker-Planck code of a three component system with stars, a fraction of dark matter in massive black holes and the rest distributed smoothly (as in the case of particle dark matter). The resulting 3 coupled partial differential equations will fully describe the evolution of all three components over time. 

In summary, we have shown that the light profile of dwarf galaxies can be used to constrain the abundance of stellar-mass black holes a the dark matter. We used Segue 1 as a generic example to demonstrate the effects of relaxation and mass segregation. Our main results are: (i) mass segregation in dwarf galaxies leads to the depletion of stars in the central regions of dwarf galaxies, and the projected stellar surface density develops a ring of higher stellar density; (ii) Segue 1 data implies that black hole dark matter fractions greater than  (6\%, 20\%) with $\mBH = (30 \Msun, 10\Msun)$ are excluded at the 99.9\% level. If future observations of dwarf galaxies show the presence of a ring in the projected stellar surface density then it will be possible to infer the fraction of dark matter made of heavy black holes with implications on primordial black holes, early universe cosmology and inflation. 

We acknowledge useful discussions with Robert Fisher, Alex Geringer-Sameth, Kyriakos Vattis and Matthew Walker. This work was supported by the Black Hole Initiative, which is funded by a grant from the John Templeton Foundation. SMK is supported by NSF PHYS-1417505 and by the Institute for Theory and Computation at the Harvard-Smithsonian Center for Astrophysics where part of this work was completed.  

\bibliography{manuscript}

\end{document}